\documentclass[aps,pre,twocolumn,superscriptaddress,showkeys,groupedaddress]{revtex4-2}

\usepackage{graphics,graphicx}

\usepackage{amsmath,amssymb,amsfonts}
\usepackage{xcolor}
\usepackage{float}
\usepackage{soul}
\usepackage[toc,page]{appendix}
\usepackage{multirow}
\usepackage[section]{placeins}
\usepackage{courier,ulem} 
\usepackage{epstopdf}
\AppendGraphicsExtensions{.tiff}

\begin{document}

\title{The Effect of Intra- and Inter-ring Couplings in Leaky Integrate-and-Fire Multiplex Networks} 

\author{K.~Anesiadis}
\affiliation{Institute of Nanoscience and Nanotechnology, National Center for Scientific Research ``Demokritos'', GR-15341, Athens, Greece}
\affiliation{School of Mathematical and Physical Sciences, National Technical University of Athens, GR-15780, Athens, Greece} 

\author{J.~Hizanidis}
\affiliation{Department of Physics, University of Crete, GR-71003, Herakleio, Greece}
\affiliation{Institute of Applied and Computational Mathematics, Foundation for Research and Technology -- Hellas, GR-70013, Herakleio, Greece}

\renewcommand{\baselinestretch}{1.5}

\begin{abstract} 
{We study the dynamics of identical Leaky Integrate-and-Fire (LIF) neurons on a multiplex composed of two ring networks with symmetric nonlocal coupling within each ring and one-to-one connections between rings. We investigate the impact of different intra-ring coupling strengths in the two rings for attractive and repulsive inter-ring coupling and show that they can lead to subthreshold oscillations. The corresponding parameter spaces where this phenomenon occurs are determined numerically. Moreover, we show that depending on whether the couplings between the two rings are attractive or repulsive, the interaction produces qualitatively different behavior in the synchronization patterns and the mean frequency profiles.}
\end{abstract}

\maketitle

\normalsize
%
\section{Introduction}

Numerical studies of networks of coupled nonlinear oscillators show interesting, rather unexpected collective patterns. 
Synchronization, which results from the both cooperative and competitive activity of the elements of the network in question, has important applications in different fields, such as physics, engineering, biology, and social sciences~\cite{pi01,st03,bo18}. 
Indicatively, the synchronization is relevant in collective dynamics of coupled physical systems (pendula, lasers, electrical oscillators), in vibration control in bridges, in power grids regarding the rotational phases of the generators, in parallel computing where the core idea is the efficient simultaneous execution of multiple actions and in vital processes related to the human physiology and the brain functionality~\cite{ka14,ni12,sh14,fu14,ro12,ci21,sc98,ax06}. 

Chimera state is a notable pattern of collective behavior in coupled nonlinear oscillators, with its identification and naming being attributed to~\cite{ku02,ab04}, where some connected parts of the system are coherent while others are not. 
In addition to the Kuramoto phase oscillator, they have been reported for several other nonlinear oscillators, including the Leaky Integrate-and-Fire (LIF)~\cite{ol10}, the FitzHugh-Nagumo (FHN)~\cite{om13,om15}, and the Hindmarsh-Rose (HR)~\cite{hi14,hi16} neuronal models. 
Since the work of Abrams \& Strogatz, significant progress has been made in expanding our understanding of the conditions under which chimera states are formed, notably for systems of coupled phase oscillators \cite{ab04,wo11,om18,om22}. 
Especially for the LIF model, another curious phenomenon is the coexistence of system elements performing complete oscillation cycles and elements that are fluctuating below the threshold potential, without surpassing it. 
These oscillations, called subthreshold~\cite{ts17}, belong to the more general category of bump states~\cite{la20} and have been reported in the field of neurophysiology in biological neurons and neural circuits~\cite{al89}. 

Network coupling topologies used in studies are mainly regular, such as global, symmetric nonlocal, small-world, and scale-free~\cite{lu10,ts16,sh17,ts21,zh14}. 
Although real-world networks are generally more complex, regular connectivities are often the starting points for understanding the underlying mechanism that drives synchronization in the more complicated, realistic systems. 
Additionally, they are more straightforward, easier to test and compare, and do not have a large computational cost. 

LIF networks of different connectivity schemes have been investigated and showed a variety of synchronization patterns. 
In networks of two populations with global connectivity, typical and periodic breathing chimera states are reported, where the ratios of coherence/incoherence for the two populations change 
periodically and in antiphase~\cite{ol10,ol19}. 
Regarding ring networks, using symmetric nonlocal connectivity the system exhibits classic and multi-headed chimera states, bump states, and solitary states~\cite{ts16,ts17}. 
Solitary states are weak chimeras where there is complete in-phase synchronization except for some relatively small clusters of incoherent oscillators. 
The term ``weak chimeras'' refers to chimera states whose incoherent domains are of infinitesimal size, while typical chimeras have both coherent and incoherent domains of finite sizes~\cite{as15}. 

This study extends the work done for single-ring networks of LIF neurons to a multilayered network (multiplex) of two structurally identical ring networks by showing the interplay of the intra-ring coupling strengths. 
Each ring of the multiplex consists of symmetric nonlocally coupled LIF neuronal oscillators while between them there are one-to-one connections. 
This two-layered network draws motivation from the composition of the brain into two hemispheres, where each node represents a brain region as identified by parcellation studies~\cite{ro15,ar18,al21}. 
We examine a parameter range for the different coupling strengths within each ring and two cases for the coupling strength between the two rings, in which significant differences were observed in the calculated time averages of the Kuramoto order parameter of the two rings. 
We show graphically how differences in intra-ring coupling strengths affect the phase synchronization in the two layers of the multiplex, as well as what is the effect of the multiplexing. 
For sufficiently large differences in intra-ring coupling strengths, we identify that one of the two rings performs complete oscillations below the threshold while the other does not.

In Sec.~\ref{sec:2} we introduce the single LIF model, the coupled LIF model, the Kuramoto order parameter, and Pearson's correlation function. 
In Sec.~\ref{sec:3} we show the triangular heatmaps of the Kuramoto order parameters for the two rings of the multiplex, independently, for the two levels of inter-ring coupling strength. 
We detect the effect of the multiplexing by highlighting the area in the maps where there is a distinct difference from the benchmark (single-ring network). 
In addition, we present a different type of subthreshold oscillations observed for specific parameter values. 
For a particular range of the control parameters, either the elements of one ring perform full-cycle oscillations while the elements of the other do not, or they even switch roles irregularly in time. 
We conclude, in Sec.~\ref{sec:4}, by summarizing our results and presenting open problems. 
%
\section{The model}\label{sec:2}

First proposed in 1907 by Louis Lapicque \cite{la07,br07}, the LIF model provides a simplified approach to the dynamics of isolated neurons. 
If we let $u(t)$ to be the time-dependent membrane potential of a nerve cell, the dynamics of the single LIF model is described by the following Eq.~\ref{eq:11} and Condition \ref{eq:12}:
\begin{subequations}
\begin{equation}\label{eq:11} 
\frac{du(t)}{dt} = \mu - u(t) + I(t) \> ,
\end{equation} 
\begin{equation}\label{eq:12} 
\lim\limits_{\delta t \to 0^+} u(t + \delta t) = u_{\rm rest} \> , \quad \mbox{when} \>\>\> u(t) > u_{\rm th} \> .
\end{equation} 
\end{subequations}
Equation~\ref{eq:11} represents the integration of the membrane potential, while influx $I(t)$ may originate from external stimuli or the neighboring neurons' collective contribution. 
Condition \ref{eq:12} represents the resetting of the potential after reaching the threshold $u_{\rm th}$. 
Namely, the potential $u(t)$ is reset at $u_{\rm rest}$ immediately after its value surpasses the value of $u_{\rm th}$. 
The parameter $\mu$ in Eq.~\ref{eq:11} corresponds to the limiting value of the potential if resetting is not considered. 
Equation~\ref{eq:11} can be analytically solved, when $I(t)$ is constant or zero. 
Then, the constant is incorporated in the parameter $\mu$ and the solution is $u(t) = \mu - (\mu - u_{\rm rest}) e^{-t}$, for $u_{\rm rest} \leq u(t) \leq u_{\rm th}$. The period $T_{\rm s}$ of oscillations of the single LIF is 
calculated as $T_{\rm s} = \ln \left[ \left( \mu - u_{\rm rest} \right) / \left( \mu 
- u_{\rm th} \right) \right]$. 
Typically, a LIF neuron spends a period of time (called refractory period) at the rest state after reset, but for simplicity, we will omit it. 
%
\subsection{The coupled LIF dynamics in the multiplex}\label{sec:21}

There are studies where LIF dynamics are demonstrated on a ring network, showing various synchronization patterns depending on the connectivity (nonlocal, hierarchical, reflecting, small-world, etc), the coupling strength, the coupling range and the refractory period \cite{ol10,ts17,ts16,ts21}. 
In this study, we consider a two-layered multiplex, where each layer is a ring network of identical LIF oscillators. Both rings are considered identical, except for the strength of the connections within each one. 
For convenience, we name them ring~${\rm L}$ (for left) and ring~${\rm R}$ 
(for right), respectively. 
For reasons of simplicity, we consider typical symmetric nonlocal connectivity within each ring and one-to-one connectivity across rings. 

\begin{figure*}[!ht]
        \centering
        \includegraphics[width=0.75\textwidth]{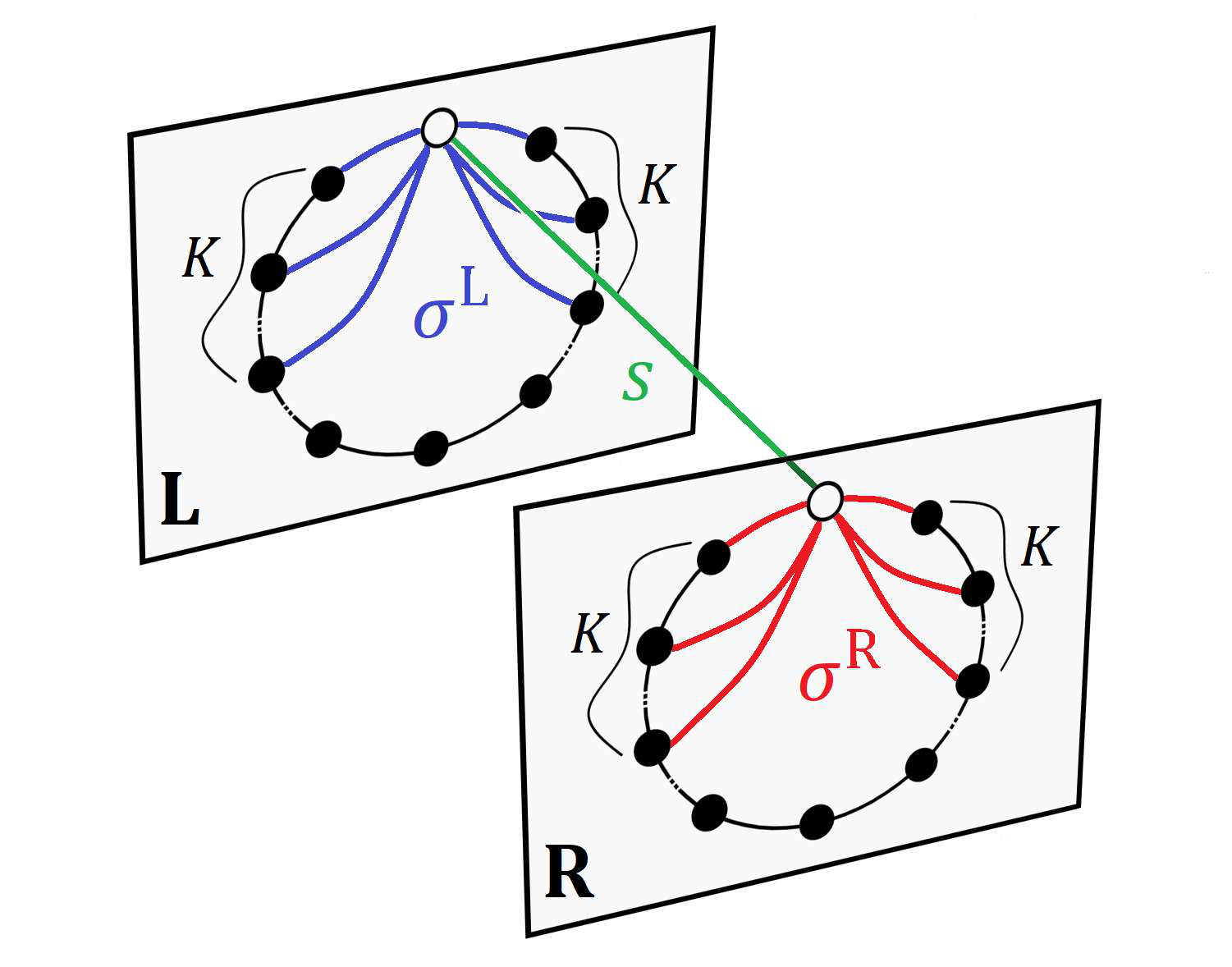}
        \caption{\small Sketch of the multiplex connectivity. The nodes in rings~L and~R are linked with inter-ring coupling strength $s$ while the nonlocal intra-ring couplings have strengths $\sigma^{\rm L}$ and $\sigma^{\rm R}$. For the sake of clarity, the connectivity of one node of ring~L within the ring ($2K$ blue junctions) and with its homologous in ring~R (green junction) is depicted, as well the connectivity of this node in ring~R within the ring ($2K$ red junctions). } 
        \label{fig:multiplex}
\end{figure*}

Let us denote by $\sigma_{jk}^{\rm L}$ the intra-ring coupling strength between nodes $(j,k)$ in ring ${\rm L}$; similarly for ring~${\rm R}$. 
To avoid having many different parameters, we assume that the connections within each layer are tantamount. 
Thus, the general form of the nonlocal intra-ring connectivity with coupling range $K$ around node $j$ of the ring~${\rm L}$ is:
\begin{equation}\label{eq:sigma} 
\sigma_{jk}^{\rm L} \equiv \sigma_{kj}^{\rm L} = \left\{ \begin{array}{ll} \sigma^{\rm L}, \quad & \forall k: \left[ j-K \leq k \leq j+K 
\right] \\ 0, & \mbox{elsewhere\>.} \end{array} \right. \end{equation} 
Similarly for ring~${\rm R}$. 
Regarding the inter-ring connections, let us denote by $\sigma_{j}^{\rm L \to R}$ the coupling strength between the $j$-th nodes of rings~${\rm L}$ and~${\rm R}$, and, similarly, for the opposite direction (note that these are the only connections between the two rings if any). As before, for the sake of simplicity, we assume common values for all connections, $\sigma_j^{\rm L \to R} \equiv \sigma_j^{\rm R \to L} \equiv s$. 
A schematic diagram of the network is shown in Fig.~\ref{fig:multiplex}. 

Let $u_j^{\rm L}(t)$, $j=1,\ldots,N$ represent the membrane potential of the $j$-th neuron in ring~L, where $N$ is the ring size. Then, the dynamics of the $j$-th coupled LIF neuron of the ring~${\rm L}$ in the multiplex is described as follows: 
\begin{subequations}\label{eq:2} 
\begin{align}\label{eq:21} 
\frac{du_j^{\rm L}(t)}{dt} &= 
    \begin{aligned}[t] &\mu - u_j^{\rm L}(t) \\ &+ \frac{\sigma^{\rm L}}{2K} \sum_{k=j-K}^{j+K} \left[ u_k^{\rm L}(t) - u_j^{\rm L}(t) \right] \\ &+ s \left[ u_j^{\rm R}(t) - u_j^{\rm L}(t) \right] \> , \end{aligned} 
\end{align} 
\begin{equation} 
\label{eq:22} \lim\limits_{\delta t \to 0^+} u_j^{\rm L}(t + \delta t) = u_{\rm rest} \> , \quad 
\mbox{when} \>\>\> u_j^{\rm L}(t) > u_{\rm th} \> . 
\end{equation} 
\end{subequations}
The notation and definitions are similar for ring ${\rm R}$. 
In Eq.~\ref{eq:2}, we consider nonlocal diffusive-like connectivity with coupling range $K$, common in both rings. 
Connections with positive coupling strength are also called attractive, while connections with negative coupling strength are also called repulsive. 
In the above expressions all the indices in the rings ${\rm L}$ and ${\rm R}$ are taken modulo $N$. 
Other common parameters of all nodes are the limiting membrane potential value $\mu$, the rest state potential $u_{\rm rest}$ and the threshold potential $u_{\rm th}$. 

In this study we use as working parameter set: $\mu = 1$, $u_{\rm rest} = 0$, $u_{\rm th} = 0.98$, $N=500$ and $K = 120$. 
For these parameters, the single (uncoupled) ring network presents subthreshold oscillations for positive coupling strength, solitary state for $\sigma \in \left[ -0.5 , 0 \right)$, chimera state for $\sigma \in \left[ -0.9 , -0.6 \right)$, incoherent state for $\sigma\simeq-1$ and again chimera state for $\sigma \in \left[ -2.0 , -1.6 \right)$~\cite{ts17,ts16}. 
In single-ring LIF networks, chimera states were reported for moderate repulsive coupling until they collapse in full disorder by further strengthening it. 
The critical value where the transition from solitary to chimera state depends on the parameter set of the LIF oscillator and the network. 
The two levels of inter-ring coupling $s$ in the multiplex connectivity we consider are $s=+0.1$ and $-0.1$. 
All simulations start from random initial conditions, while periodic boundary conditions are considered for all indices. 
%
\subsection{Kuramoto order parameter and intercorrelation}\label{sec:22}

For quantifying the synchronization within each ring the Kuramoto order parameter $Z$ is employed \cite{ku02,om13}, denoted by $Z^{\rm L}$ and $Z^{\rm R}$ for the rings ${\rm L}$ and ${\rm R}$ respectively. 
To define $Z$ we first need to define the phase for every oscillator. 
Then, the instantaneous Kuramoto order parameter which measures synchronization in ring ${\rm L}$ is defined as:
\begin{equation} \label{eq:kura} 
Z^{\rm L}(t) = \frac{1}{N^{\rm L}} \left| \sum_{k=1}^{N} e^{i \phi_k^{\rm L}(t)} \right| \> ,
\end{equation} 
where $\left| \cdot \right|$ stands for the magnitude of the complex number in the argument. 
Similarly, the Kuramoto order parameter $Z^{\rm R}(t)$ is defined for ring ${\rm R}$. 
The order parameter takes values in the range $0 \leq Z(t) \leq 1$. When $Z \simeq 0$ then the ring elements are incoherent and when $Z \simeq 1$ they are synchronized in phase. 
Intermediate values of $Z$ indicate partial network synchronization. 
Typically, solitary states exhibit almost absolute coherence with the incoherent oscillators consisting only of a small fraction of the network and thus $Z$ is very close to $1$. 
On the other hand, typical chimeras have a finite domain of incoherence, reflected in the Kuramoto order parameter taking values considerably less than $1$. 

On the definition of the phase, in this study the instantaneous phase $\phi_j^{\rm L}(t)$  of the $j$-th oscillator in ring ${\rm L}$ is given by: 
\begin{equation}\label{eq:phase} 
\phi_j^{\rm L}(t) = \frac{2 \pi u_j^{\rm L}}{u_{\rm th}} \> . 
\end{equation} 
Similarly, are defined the phases in the ring ${\rm R}$. 
This definition along with the definition used in \cite{ol10} are found to produce consistent results. 

To quantify the inter-synchrony between the two rings, we employ the linear Pearson's correlation function $C^{\rm L-R}$~\cite{ni15}, defined as 
\begin{equation}\label{eq:corr}
    C^{\rm L-R}(t) = \frac{\left< u_j^{\rm L} u_j^{\rm R} \right> - \left< u_j^{\rm L} \right> \left< u_j^{\rm R}\right>}{ \left( \left< u_j^{\rm L} u_j^{\rm L} \right> - \left< u_j^{\rm L} \right>^2 \right) \left( \left< u_j^{\rm R} u_j^{\rm R} \right> - \left< u_j^{\rm R} \right>^2 \right)} \> ,
\end{equation} 
where the averages $\left< \cdot \right>$ are defined over all $j = 1, \ldots, N$ oscillators in each ring. 
When $\left| C^{\rm L-R} \right| \to 1 $, the measure indicates full inter-synchrony, while vanishing values indicate the absence of (linear) correlation between the two rings. 
We will refer to $C^{\rm L-R}$ also as intercorrelation. 
%
\section{Results}\label{sec:3} 

For the above system and the working parameter set at the end of Sec.~\ref{sec:21}, we present our results on our performed numerical simulations for a grid of $0.1\times0.1$ boxes showing the magnitudes of intra-ring coupling strength values $\sigma^{\rm L}$ and $\sigma^{\rm R}$, ranging from $-2.0$ to $+1.0$ and for inter-ring coupling strength $s=+0.1$ and $s=-0.1$. 
Roughly speaking, this intensity of interaction represents relatively weak attractive and repulsive interaction between the two rings, respectively. 
Due to the resolution of the grid that is used, there may be other stable states of the multiplex that are overlooked for some off-grid parameter tuples. 
Throughout this section, the intra-ring coupling strengths are interchangeable as the two rings are in everything else identical. 

\begin{figure*}[!ht]
        \centering
        \includegraphics[width=1.00\textwidth]{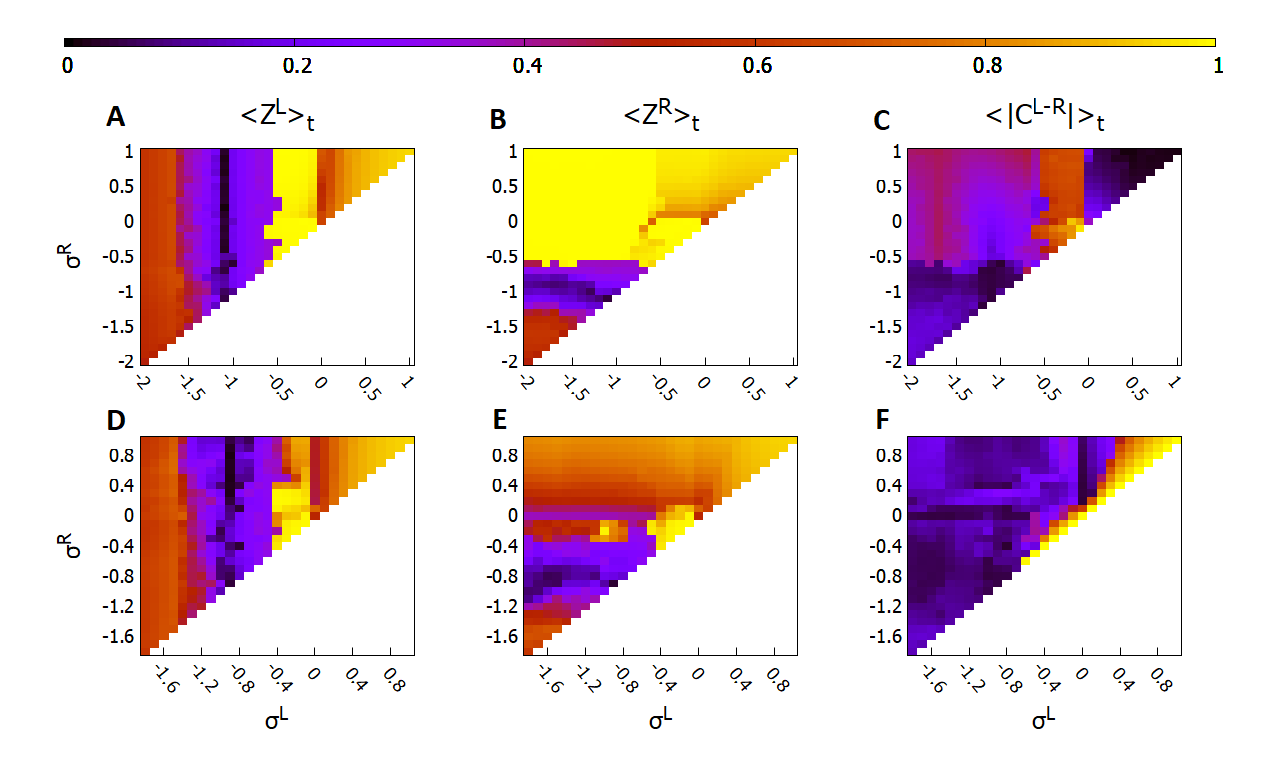}
        \caption{Color-maps of the magnitude of time-averaged Kuramoto order parameters $\left< Z^{\rm L}\right>_t$, $\left< Z^{\rm R}\right>_t$ of the multiplex and its time-averaged absolute intercorrelation $\left< \left| C^{\rm L-R} \right| \right>_t$ for $s=+0.1$ ({\bf A}, {\bf B} and {\bf C}, respectively) and for $s=-0.1$ ({\bf D}, {\bf E} and {\bf F}, respectively), versus intra-ring coupling strength $\sigma^{\rm L}$ and $\sigma^{\rm R}$. 
        For each value in the heat maps, the evolution 
        time of the system was $4000$ Time Units (TUs). 
        The initial $2000$ TUs were discarded as transient. 
        Other parameters are: $N=500$, $K=120$, $\mu=1$, $u_{\rm rest}=0$ and 
        $u_{\rm th}=0.98$. 
        All simulations were performed starting from the same random initial 
        conditions.} 
        \label{fig:maps}
\end{figure*}

We show in Fig.~\ref{fig:maps} the heatmaps for the time-averaged of the Kuramoto order parameters $\left< Z^{\rm L}\right>_t$, $\left< Z^{\rm R}\right>_t$ and time-averaged absolute correlation $\left< \left| C^{\rm L-R} \right| \right>_t$ between the two rings, for attractive coupling $s=+0.1$ and for repulsive coupling $s=-0.1$. 
On the diagonals, intra-ring coupling strengths coincide, $\sigma^{\rm L} \equiv \sigma^{\rm R}$, while on the upper triangular parts $\sigma^{\rm L} < \sigma^{\rm R}$ (the lower triangular part of the Figs.~\ref{fig:maps}A,~D is actually mirrored on Figs.~\ref{fig:maps}B,~E, respectively, and vice versa). 
In this study, the effect of one ring on the other is examined. 
Therefore, we focus on the color differentiation happening above the color boxes of the diagonals of Figs.~\ref{fig:maps}A,~D -- we fix $\sigma^L$ and vary $\sigma^R$. 
The simulations were done using the same set of randomized initial conditions. 
Different initial conditions may slightly shift the major transitions where they occur on the colormaps e.g. for high phase synchronization to low. The scenario where $\sigma^{\rm L}=\sigma^{\rm R}$ has been reported in previous work~\cite{an22}. 

\begin{figure*}[!ht]
        \centering
        \includegraphics[width=0.90\textwidth]{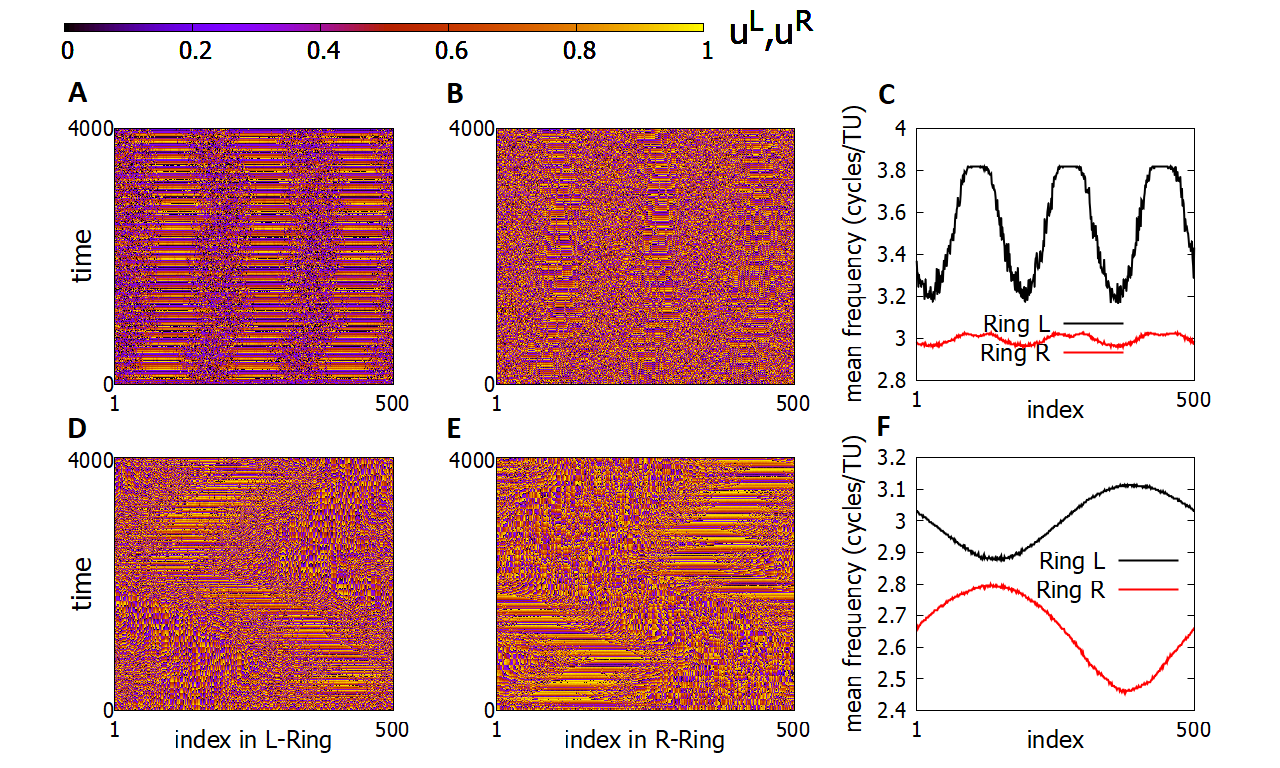} 
        \caption{\small {\bf Top}: Spacetime plots of the membrane potentials for the rings~L~and~R on {\bf A} and {\bf B}, respectively, and their frequency profiles on {\bf C} when $\sigma^{\rm L}=-1.8$ and $\sigma^{\rm R}=-0.9$. \\
        {\bf Bottom}: Spacetime plots of the membrane potentials for the rings~L~and~R on {\bf D} and {\bf E}, respectively, and their frequency profiles on {\bf F} when $\sigma^{\rm L}=-0.9$ and $\sigma^{\rm R}=-0.7$. \\ 
        The intra-ring coupling strength is $s=+0.1$. 
        In both cases, the evolution time of the system was 4000 TUs of which the initial 2000 TUs were discarded as transient time.  
        Other parameters of the multiplex are as in Fig.~\ref{fig:maps}. 
        All simulations start from the same random initial conditions. }
        \label{fig:sp1}
\end{figure*}

\begin{figure*}[!ht]
        \centering
        \includegraphics[width=0.90\textwidth]{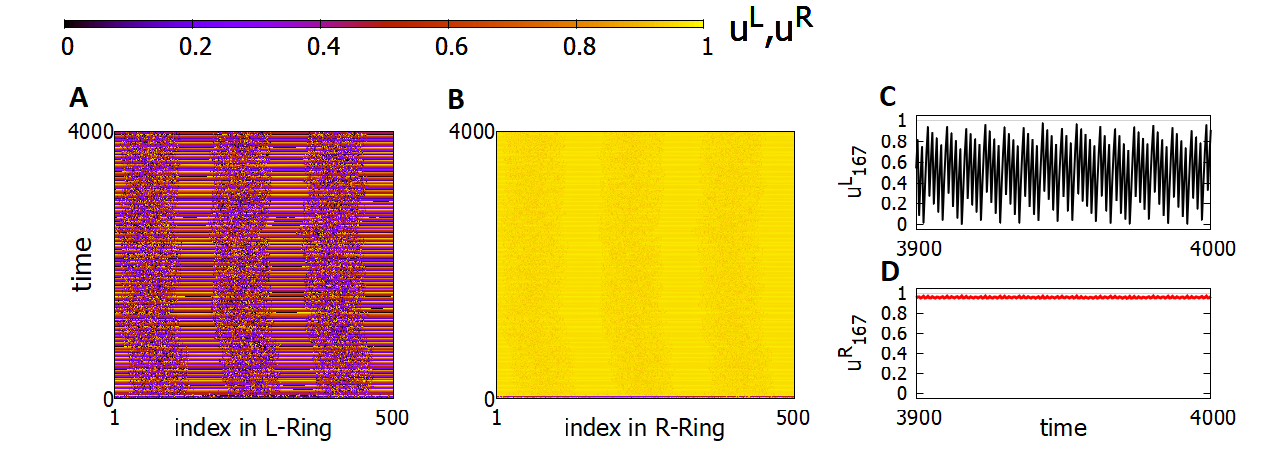}
        \caption{\small Spacetime plots of the membrane potentials for the rings~L~and~R on {\bf A} and {\bf B}, respectively, and time-series of $u^{\rm L}_{167}$,~$u^{\rm R}_{167}$ on {\bf C} and {\bf D} when $s=+0.1$, $\sigma^{\rm L}=-1.7$ and $\sigma^{\rm R}=-0.5$. \\
        Other parameters of the multiplex are as in Fig.~\ref{fig:maps}. 
        All simulations start from the same random initial conditions. }
        \label{fig:sp2}
\end{figure*}

\begin{figure*}[!ht]
        \centering
        \includegraphics[width=0.85\textwidth]{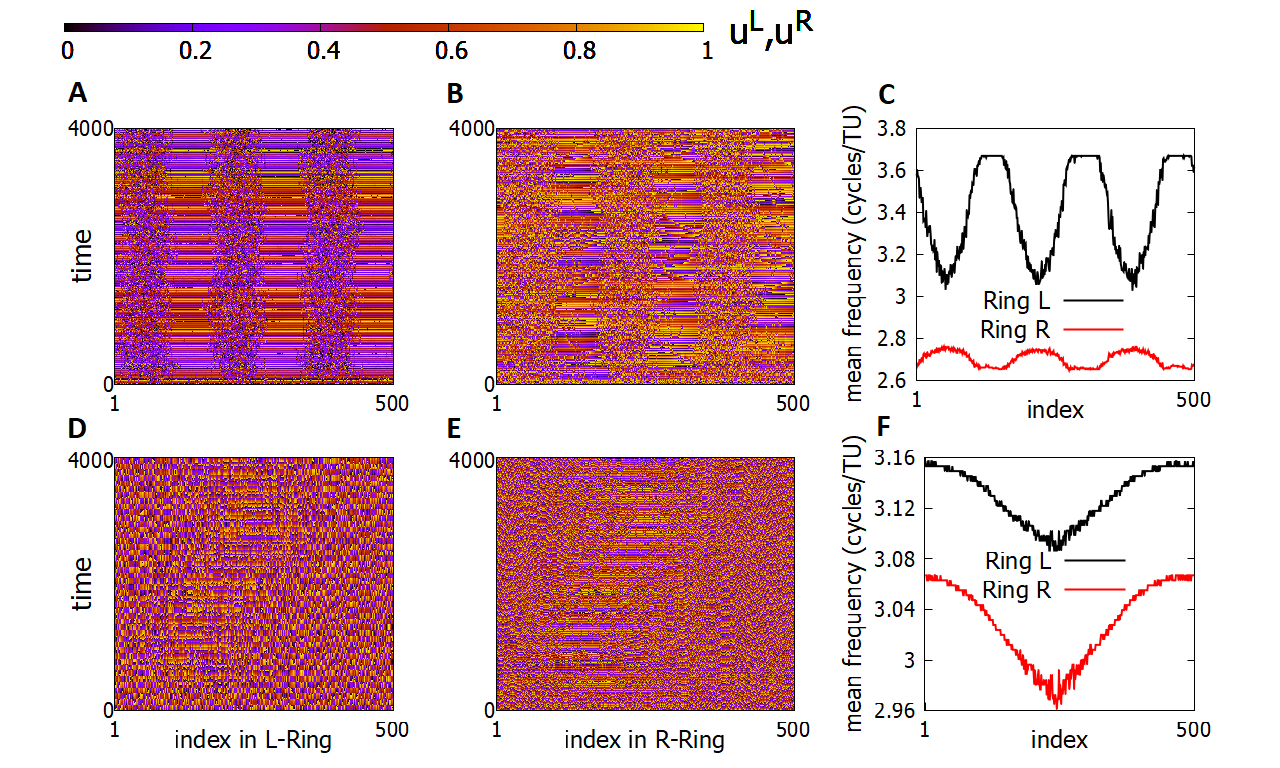}
        \caption{\small {\bf Top}: Spacetime plots of the membrane potentials for the rings~L~and~R on {\bf A} and {\bf B}, respectively, and their frequency profiles on {\bf C} when $\sigma^{\rm L}=-1.6$ and $\sigma^{\rm R}=-0.4$. \\
        {\bf Bottom}: Spacetime plots of the membrane potentials for the rings~L~and~R on {\bf D} and {\bf E}, respectively, and their frequency profiles on {\bf F} when $\sigma^{\rm L}=-0.8$ and $\sigma^{\rm R}=-0.7$. \\ 
        The intra-ring coupling strength is $s=-0.1$. 
        In both cases, the evolution time of the system was 4000 TUs of which the initial 2000 TUs were discarded as transient time.  
        Other parameters of the multiplex are as in Fig.~\ref{fig:maps}. 
        All simulations start from the same random initial conditions. }
        \label{fig:sp3}
\end{figure*}

\begin{figure*}[!ht]
        \centering
        \includegraphics[width=0.85\textwidth]{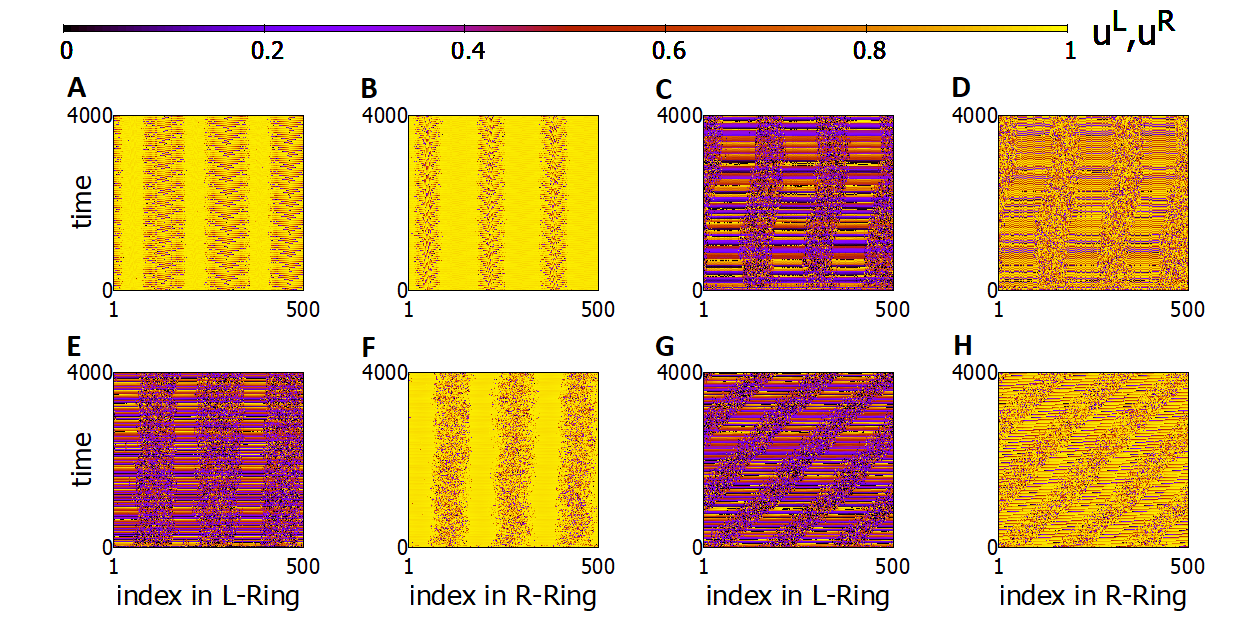}
        \caption{\small Spacetime plots of the membrane potentials for rings~L~and~R. \\
        {\bf Top}: {\bf A}, {\bf B} $\sigma^{\rm L}=+0.2$, $\sigma^{\rm R}=+0.4$, $s=+0.1$; $\quad$ {\bf C}, {\bf D} $\sigma^{\rm L}=-1.6$, $\sigma^{\rm R}=+0.2$, $s=-0.1$. \\
        {\bf Bottom}: {\bf E}, {\bf F} $\sigma^{\rm L}=-1.7$, $\sigma^{\rm R}=+0.9$, $s=-0.1$; $\quad$ {\bf G}, {\bf H} $\sigma^{\rm L}=-1.8$, $\sigma^{\rm R}=+0.4$, $s=-0.1$. \\
        In both cases, the evolution time of the system was 4000 TUs of which the initial 2000 TUs were discarded as transient time.  
        Other parameters of the multiplex are as in Fig.~\ref{fig:maps}. 
        All simulations start from the same random initial conditions. }
        \label{fig:sp4}
\end{figure*}

Either for $s=+0.1$ or $-0.1$, the Kuramoto order parameter $Z^{\rm L}$ of the ring~L is generally consistent as $\sigma^{\rm R}$ varies. 
From $\sigma^{\rm L}=-2.0$ to $-1.4$, $Z^{\rm L}$ takes intermediate values, indicating coherence and chimera states, while from $\sigma^{\rm L}=1.3$ to $-1.0$, $Z^{\rm L}$ tends to 0 and ring~L gets fully incoherent. 
Notice that for $s=+0.1$, incoherence occurs near $\sigma^{\rm L}=-1.1$ when for $s=-0.1$ it occurs near $\sigma^{\rm L}=-0.9$. 
For even smaller sizes of $\sigma^{\rm L}$, $Z^{\rm L}$ increases monotonously towards 1, pointing out that the oscillating activity is almost in perfect synchrony. 
Turning to ring~R and for $-0.5 \leq \sigma^{\rm R} \leq 0$, the Kuramoto order parameter $Z^{\rm R}$ takes values close to 1 when $s=+0.1$, but intermediate values when $s=-0.1$. 
In the former scenario, where $Z^{\rm R} \simeq 1$, it appears to be insensitive to the $\sigma^{\rm L}$ (except for the $\sigma^{\rm L}=-0.5$). 
However, as we show below, a transition in the qualitative behavior of ring~R takes place which coincidentally does not alter $Z^{\rm R}$, i.e. the level of in-phase synchrony. 
As for the calculated averages of the correlation between the two rings, they are relatively high for $s=+0.1$, $-2.0 \leq \sigma^{\rm L} \leq -1.4$ and $-0.5 \leq \sigma^{\rm R} < 0$. 
The intercorrelation is maximized for low values of intra-ring coupling strengths, where they are comparable to the inter-ring coupling strength ($|s|=0.1$). 
In addition, for the most part, it loosens wherever at least one of the rings shows vanishing $Z$ values or for solely attractive connections; see Figs.~\ref{fig:maps}C,~F. 

We highlight our observations concerning attractive weak multiplexing $s=+0.1$ and repulsive coupling within each ring. 
In the range $-2.0 \leq \sigma^{\rm L} \leq -1.6$, most of the time ring~L develops a relatively localized three-headed chimera state -- that is the simultaneous presence of three coherent and three incoherent clusters of oscillators -- actuating ring~R to do also even for coupling strength values which are far from giving coherent clusters that persist in time. 
An example is shown in Fig.~\ref{fig:sp1}A,B for $\sigma^{\rm L}=-1.8$ and $\sigma^{\rm R}=-0.9$. 
The mean frequencies in both rings are maximized towards the centers of the coherent clusters while forming plateaus on the incoherent ones, as shown in Fig.~\ref{fig:sp1}C. 
Lowering the $-1.5 \leq \sigma^{\rm L} < -1.2$, the chimera states in ring~L give way to synchronized oscillations that are made temporarily and sparsely in the network to eventually dissipate. 
The activity in ring~L, resembling the frequency profile in Fig.~\ref{fig:sp1}C, inhibits ring~R to develop typical chimera states for $-0.8 \leq \sigma^{\rm R} \leq -0.6$, as we would expect. 
 However, the frequency profile is flattened for $-1.2 \leq \sigma^{\rm L} < -1$, where ring~L enters full disorder. 
In the range $-1.0 \leq \sigma^{\rm L} < -0.7$, ring~L exhibits primarily chimera states drifting randomly in time, as shown in Fig.~\ref{fig:sp1}D for $\sigma^{\rm L}=-0.9$ and $\sigma^{\rm R}=-0.7$. 
Using this case as an example, the two chimeras, shown in Figs.~\ref{fig:sp1}D,~E, are positioned in opposite sites in the two rings. 
Another difference from the case shown in Figs.~\ref{fig:sp1}A,~B,~C is that the mean frequencies peak at the centers of the incoherent regions of the respective rings. 
Finally, for $-0.7 < \sigma^{\rm L} < 0$, both rings exhibit mainly solitary states. 

In Fig.~\ref{fig:maps}B we notice for $\left( \sigma^{\rm L}, \sigma^{\rm R} \right) \in \left[ -2.0, -0.8\right]\times\left[ -0.5, 0\right)$ an abrupt change of $Z^{\rm R}$. 
In this parameter area, all the oscillators in the ring~R -- with the weakest intra-ring coupling -- from the beginning of the calculations are almost immediately set to perform non-complete oscillations below the threshold potential. 
Furthermore, they imitate the oscillating activity in ring~L; e.g. for $\sigma^{\rm L}=-1.7$, $\sigma^{\rm R}=-0.5$ we show in Figs.~\ref{fig:sp2}A,~B that the ring~R imitates the three-headed chimera of the ring~L. 
Indicatively, the time-series of $u^{\rm L}_{167}$ (plotted in Fig.~\ref{fig:sp2}C) covers the entire range $\left[ 0 , u_{\rm th} \right]$ while $u^{\rm R}_{167}$ fluctuates very close to the threshold $u_{\rm th}$. 

Concerning repulsive coupling between the two rings for $s=-0.1$ and repulsive coupling within each ring, for large values of intra-ring coupling strengths $-1.8 \leq \sigma^{\rm L} < \sigma^{\rm R} \leq -1.4$ the mean frequency profiles are rather noisy and no chimera states are clearly formed. 
For smaller values of $\sigma^{\rm R}$, three-headed chimera states appear in ring~L and its mean frequency profile smoothens out and forms plateaus on the coherent domains and arcs on the incoherent ones. 
In particular, for $-0.4 \leq \sigma^{\rm R} < 0$, three-headed chimeras are also formed in ring~R. 
We show a such case in Fig.~\ref{fig:sp3}A and~B for $\sigma^{\rm L}=-1.6$ and $\sigma^{\rm R}=-0.4$. 
The coherent domains in one and the other ring coincide in the same locations (similarly for the incoherent domains). 
Notice that in Fig.~\ref{fig:sp3}C, for ring~L the ``valleys" correspond to incoherent oscillators but for ring~R they correspond to coherent oscillators; respectively for the ``hills". 
In the range $-1.4 \leq \sigma^{\rm L} < -1.0$, the multiplex displays scattered short-time synchronized clusters, even in ring~R for low values of $\sigma^{\rm R}$ for which we would expect a state of high organization, such as a solitary state. 
This is also indicated by checking that the Kuramoto order parameters in the corresponding part of Fig.~\ref{fig:maps}E compared to its respective in Fig.~\ref{fig:maps}B are generally lower. 
Solitary states are only observed for $-0.5 < \sigma^{\rm L},\sigma^{\rm R} < 0$. 
Typical one-headed chimera states are observed for $-1.0 \leq \sigma^{\rm L} \leq -0.6$. 
Indicatively, we show in Fig.~\ref{fig:sp3}D,~E and~F for $\sigma^{\rm L}=-0.8$, $\sigma^{\rm R}=-0.7$ that both rings develop chimera states with their characteristic mean frequency profiles. 

Concluding the results section we consider attractive intra-ring coupling, starting with the case $s=+0.1$. 
For $-2.0 \leq \sigma^{\rm L} < -0.6$ and $0 < \sigma^{\rm R}$, the elements of ring~R perform subthreshold oscillations which resemble in pattern the oscillations in ring~L, similar to those shown in Fig.~\ref{fig:sp2} (while for $-0.6 \leq \sigma^{\rm L} < 0$, a few elements of ring~R shortly escape from this behavior). 
When we switch to positive $\sigma^{\rm L}$, we notice in Figs.~\ref{fig:maps}A,~D a drop in $Z^{\rm L}$ to $0.5$ that increases again monotonously towards 1 as $\sigma^{\rm L}$ increases. 
We report the coexistence of subthreshold oscillatory domains and domains of moderate coherence, as in Fig.~\ref{fig:sp4}A, B. 
In addition, the multiplex shows low intercorrelation, where the subthreshold oscillatory domains in one ring are opposite mid-coherent domains in the other ring and vice versa. 
The ring~R, with the larger value of coupling strength, exhibits more subthreshold oscillations than ring~L. 
Regarding $s=-0.1$, attractive coupling in both rings induces subthreshold oscillations and full oscillations that travel across the ring. 
Finally, we report some patterns observed for $-2.0 \leq \sigma^{\rm L} < -1.4$. 
In this range, ring~L develops three-headed chimera states which either are stationary or travel across the ring in time (with or without constant speed). 
In ring~R, there are highly incoherent clusters of elements, formed spatially where the incoherent clusters of the chimera states are located in the ring~L. 
Between those clusters, we observe solitary states for $0 < \sigma^{\rm R} < +0.3$, subthreshold oscillations for $\sigma^{\rm R} \geq +0.5$ (which become localized when $+0.7 \leq \sigma^{\rm R} \leq +0.9$) and again subthreshold oscillations except for a few elements performing complete cycles for $+0.2 < \sigma^{\rm R} \leq +0.4$, corresponding to the state in the other ring; see Figs.~\ref{fig:sp4}C,~D, Figs.~\ref{fig:sp4}E,~F and Figs.~\ref{fig:sp4}G,~H, respectively. 
%
\section{Conclusion}\label{sec:4} 

In the presented study we discussed the interplay of intra-ring coupling strengths $\sigma^{\rm L}, \sigma^{\rm R} \in  \left[ -2, 1 \right]$ in a multiplex of two rings of symmetrically and nonlocally coupled LIF oscillators. 
The two rings are coupled together with one-to-one couplings with strength $s$. 
The case $\sigma^{\rm L}=\sigma^{\rm R}$,~$s=\pm0.1$ is reported in a previous work \cite{an22}, so here we consider only the case where $\sigma^{\rm L}\neq\sigma^{\rm R}$. 

Outlining the results, for attractive coupling $s=+0.1$, three-headed chimera states appear in both rings for large value of $\sigma^{\rm L}, \sigma^{\rm R} \in \left[ -2.0 , -1.4 \right)$, while most of the time for $-0.6 < \sigma^{\rm R}$, all the elements in ring~R perform subthreshold fluctuations without completing any oscillation cycle. 
On the other hand, for repulsive inter-ring coupling $s=-0.1$, we observe three-headed chimera states in the two rings. 
The mean frequency profiles of these chimeras are inverted compared to those observed of the attractive inter-ring coupling. 
In the range $\sigma^{\rm L} \in \left[-1.8 , -1.4 \right)$, attractive coupling within ring~R induces a motif where incoherent domains coexist with domains of either synchronized, subthreshold oscillators or subthreshold oscillators together with a few elements that perform complete cycles. 
For intermediate values of $\sigma^{\rm L}, \sigma^{\rm R} \in \left[ -1.0 , -0.6 \right]$, we notice typical chimera states such that the coherent domain of the one ring is located in the respective locations where the coherent domain of the other ring is when $s=+0.1$, while the opposite holds true when $s=-0.1$. 
Throughout the parameter range, frequencies of the strong intra-coupled ring (ring~L) are higher than those of the other ring (ring~R) with very few exceptions. 
Finally, the (linear) correlation between the two rings, measured by Eq.~\ref{eq:corr}, vanishes where coherence is very low in at least one ring (except for solely attractive couplings) and gets relatively large where subthreshold oscillations occur. 
It becomes maximum in the range $-0.5 < \sigma^{\rm L},\sigma^{\rm R} < 0$ for positive inter-ring coupling, and in the range $+0.4 < \sigma^{\rm L}, \sigma^{\rm R} < +1.0$ for negative inter-ring coupling.  

We conclude that the activity of the ring network with the stronger repulsive coupling to some extent dictates the activity on the other ring network and that attractive and repulsive inter-ring coupling affects the system in different ways. 
In particular, attractive coupling gives rise to typical chimera states in the multiplex where the coherent domains are located at opposite sites, while repulsive coupling creates localized three-headed chimera states in the ring with a considerably low magnitude of intra-ring coupling strength. 
Moreover, attractive coupling induces subthreshold oscillations in the entire LIF ring network with the lower intra-ring coupling, unlike the case of repulsive coupling where this is not observed. 
Unexpectedly, this is a rather robust phenomenon that persists for long simulation times and has also been evidenced by the authors for a two-layered multiplex with reflecting connectivity, where instead of one-to-one connections, each element of one ring is coupled to a symmetrical neighborhood of the element counterpart in the other ring. 
In further studies on the LIF multiplex network, it will be useful to examine whether these patterns occur also under different connectivity schemes, and inhomogeneities in the network architecture between the layers, or the LIF oscillator parameters. 
%
\section*{Funding} 

This work was supported by computational time granted from the Greek Research and Technology Network (GRNET) in the National HPC facility - ARIS, under project IDs PR009012 and PR12015.
%
\section*{Acknowledgements}

The authors would like to thank Astero Provata for helpful discussions. 
%

%
%

\begin{thebibliography}{999} 

\bibitem{pi01}
Pikovsky, A., Rosenblum, M. and Kurths, J. (2001), 
{\it Synchronization -- A Universal Concept in Nonlinear Sciences}, 
Cambridge University Press: Cambridge

\bibitem{st03}  
Strogatz, S. (2003), 
{\it Sync: The Emerging Science of Spontaneous Order}, 
Penguin Books: London

\bibitem{bo18}
Boccaletti, S., Pisarchik, A.N., Del Genio, C.I. and Amann, A. (2018), 
{\it Synchronization: From Coupled Systems to Complex Networks}, 
Cambridge University Press: Cambridge 

\bibitem{ka14} 
Kapitaniak, T., Kuzma, P., Wojewoda, J., Czolczynski, K. and Maistrenko, Y. (2014), 
Imperfect chimera states for coupled pendula, 
{\it Scientific Reports}, {\bf 4}, 6379 
\url{doi:10.1038/srep06379}

\bibitem{ni12} 
Nixon, M., Fridman, M., Ronen, E., Friesem, A.~A., Davidson, N. and Kanter, I. (2012), 
Controlling Synchronization in Large Laser Networks, 
{\it Physical Review Letters}, {\bf 108}, 214101 
\url{doi:10.1103/PhysRevLett.108.214101}

\bibitem{sh14} 
Shukla, N., Parihar, A., Freeman, E., Paik, H., Stone, G., Narayanan, V., Wen, H., Cai, Z., Gopalar, V., Engel-Herbert, R., Scholm, D.~G., Raychowdhury, A. and Datta, S. (2014), 
Synchronized charge oscillations in correlated electron systems, 
{\it Scientific Reports}, {\bf 4}, 4964 
\url{doi:: 10.1038/srep04964} 

\bibitem{fu14} 
Fujino, Y. and Siringoringo, D. (2014), 
A Conceptual Review of Pedestrian-Induced Lateral Vibration and Crowd Synchronization Problem on Footbridges, 
{\it Journal of Bridge Engineering}, {\bf 21}, C4015001 
\url{doi:10.1061/(ASCE)BE.1943-5592.0000822}

\bibitem{ro12} 
Rohden, M., Sorge, A., Timme, M. and Witthaut, D. (2012), 
Self-Organized Synchronization in Decentralized Power Grids, 
{\it Physical Review Letters}, {\bf 109}, 064101 
\url{doi:10.1103/PhysRevLett.109.064101}

\bibitem{ci21} 
Cicirelli, F., Giordano, A. and Mastroianni, C. (2021), 
Analysis of Global and Local Synchronization in Parallel Computing, 
{\it IEEE Transactions on Parallel and Distributed Systems}, {\bf 32}, 988--1000 
\url{doi:10.1109/TPDS.2020.3037469} 

\bibitem{sc98} 
Sch{\"a}fer, C., Rosenblum, M., Kurths, J. and Abel, H.-H. (1998), 
Heartbeat synchronized with ventilation, 
{\it Nature}, {\bf 392}, 239--240
\url{doi:10.1038/32567} 

\bibitem{ax06} 
Axmacher, N., Mormann, F., Fern{\'a}ndez, G., Elger, C.~E. and Fell, J. (2006), 
Memory formation by neuronal synchronization, 
{\it Brain Research Reviews}, {\bf 52}, 170--182 
\url{doi:10.1016/j.brainresrev.2006.01.007}

\bibitem{ku02}
Kuramoto, Y. and Battogtokh, D. (2002), 
Coexistence of coherence and incoherence in nonlocally coupled phase oscillators, 
{\it Nonlinear Phenomena in Complex Systems}, {\bf 5}, 380--385 
\url{doi:10.48550/arXiv.cond-mat/0210694}

\bibitem{ab04}
Abrams, D.~M. and Strogatz, S.~H. (2004), 
Chimera states for coupled oscillators, 
{\it Physical Review Letters}, {\bf 93}, 174102
(2004) \url{doi:10.1103/PhysRevLett.93.174102}

\bibitem{ol10}
Olmi, S., Politi, A. and Torcini, A. (2010), 
Collective chaos in pulse-coupled neural networks, 
{\it Europhysics Letters}, {\bf 92}, 60007 
\url{doi:10.1209/0295-5075/92/60007}

\bibitem{om13}
Omelchenko, I., Omel'chenko, O. E., H{\"o}vel, P. and Sch{\"o}ll, E. (2013), 
When Nonlocal Coupling between Oscillators Becomes Stronger: Patched Synchrony or Multichimera States, 
{\it Physical Review Letters}, {\bf 110}, 224101 
\url{doi:10.1103/PhysRevLett.110.224101}

\bibitem{om15} 
Omelchenko, I., Provata, A., Hizanidis, J., Sch{\"o}ll, E. and H{\"o}vel, P. (2015), 
Robustness of chimera states for coupled FitzHugh-Nagumo oscillators, 
{\it Physical Review E}, {\bf 91}, 022917 
\url{doi:10.1103/PhysRevE.91.022917}

\bibitem{hi14} 
Hizanidis, J., Kanas, V., Bezerianos, A. and Bountis, T. (2014), 
Chimera states in networks of nonlocally coupled Hindmarsh-Rose neuron models, 
{\it International Journal of Bifurcation and Chaos}, {\bf 24}, 1450030 
\url{doi:10.1142/S0218127414500308} 

\bibitem{hi16}
Hizanidis, J., Kouvaris, N.E., Zamora-L{\'o}pez, G., D{\'i}az-Guilera and A., Antonopoulos, C.G. (2016), 
Chimera-like States in Modular Neural Networks, 
{\it Scientific Reports}, {\bf 6}, 19845 
\url{doi:10.1038/srep19845}

\bibitem{wo11} 
Wolfrum, M. and Omel'chenko, O.~E. (2011), 
Chimera states are chaotic transients, 
{\it Physical Review E}, {\bf 84}, 015201(R) 
\url{doi:10.1103/PhysRevE.84.015201} 

\bibitem{om18} 
Omel'chenko, O.~E. (2018), 
The mathematics behind chimera states, 
{\it Nonlinearity}, {\bf 31}, R121 
\url{doi:10.1088/1361-6544/aaaa07}

\bibitem{om22} 
Omel'chenko, O.~E. (2022), 
Mathematical Framework for Breathing Chimera States, 
{\it Journal of Nonlinear Science}, {\bf 32}, 22 
\url{doi:10.1007/s00332-021-09779-1}

\bibitem{ts17}
Tsigkri-DeSmedt, N.D., Hizanidis, J., Schöll, E., Hövel, P. and Provata, A. (2017), 
Chimeras in leaky integrate-and-fire neural networks: effects of reflecting connectivities, 
{\it The European Physical Journal B}, {\bf 90}, 139 
\url{doi:10.1140/epjb/e2017-80162-0}

\bibitem{la20} 
Laing, C.R. and Omel'chenko, O. (2020), 
Moving bumps in theta neuron networks, 
{\it Chaos}, {\bf 30}, 043117 
\url{doi:10.1063/1.5143261} 

\bibitem{al89} 
Alonso, A. and Llin{\'a}s, R.R. (1989), 
Subthreshold Na$^{+}$-dependent theta-like rhythmicity in stellate cells of entorhinal cortex layer II, 
{\it Nature}, {\bf 342}, 175--177 
\url{doi:10.1038/342175a0}

\bibitem{ts16}
Tsigkri-DeSmedt, N.D., Hizanidis, J., Hövel, P. and Provata, A. (2016), 
Multi-chimera states and transitions in the Leaky Integrate-and-Fire model with nonlocal and hierarchical connectivity, 
{\it The European Physical Journal Special Topics}, {\bf 225}, 1149--1164 
\url{doi:10.1140/epjst/e2016-02661-4}

\bibitem{lu10}
Luccioli S. and Politi, A. (2010), 
Irregular collective behavior of heterogeneous neural networks, 
{\it Physical Review Letters}, {\bf 105}, 158104
\url{doi:10.1103/PhysRevLett.105.158104}

\bibitem{sh17}
Shena, J., Hizanidis, J., H{\"o}vel, P. and Tsironis, G.~P. (2017), 
Multiclustered chimeras in large semiconductor laser arrays with nonlocal interactions, 
{\it Physical Review E}, {\bf 96}, 032215 
\url{doi:10.1103/PhysRevE.96.032215}

\bibitem{ts21}
Tsigkri-DeSmedt, N.~D., Sarlis, N.~V. and Provata, A. (2021), 
Shooting solitaries due to small-world connectivity in leaky integrate-and-fire networks, 
{\it Chaos}, {\bf 31}, 083129 
\url{doi:10.1063/5.0055163}

\bibitem{zh14} 
Zhu, Y., Zheng, Z. and Yang, J. (2014), 
Chimera states on complex networks, 
{\it Physical Review E}, {\bf 89}, 022914 
\url{doi:10.1103/PhysRevE.89.022914}

\bibitem{ol19} 
Olmi, S. and Torcini, A. (2019), 
Chimera states in pulse coupled neural networks: the influence of dilution and noise, 
{\it Nonlinear Dynamics in Computational Neuroscience}, Editors F. Corinto and A. Torcini (Cham, Switzerland: PoliTo Springer Series), 65-79, Chap. 5 
\url{doi:10.1007/978-3-319-71048-8_5} 

\bibitem{as15} 
Ashwin, P. and Burylko, O. (2015), 
Weak chimeras in minimal networks of coupled phase oscillators, 
{\it Chaos}, {\bf 25}, 013106 
\url{doi:10.1063/1.4905197}

\bibitem{ro15} 
Rolls, E.~T., Joliot, M. and Tzourio-Mazoyer, N. (2015), 
Implementation of a new parcellation of the orbitofrontal cortex in the automated anatomical labeling atlas, 
{\it NeuroImage}, {\bf 122}, 1--5 
\url{doi:10.1016/j.neuroimage.2015.07.075}

\bibitem{ar18} 
Arslan, S., Ktena, S.~I., Makropoulos, A., Robinson, E.~C., Rueckert, D. and Parisot, S. (2018), 
Human brain mapping: A systematic comparison of parcellation methods for the human cerebral cortex, 
{\it NeuroImage}, {\bf 170}, 5--30 
\url{doi:10.1016/j.neuroimage.2017.04.014} 

\bibitem{al21} 
Albers, K.~J., Ambrosen, K.~S., Liptrot, M.~G., Dyrby, T.~B., Schmidt, M.~N. and M{\/o}rup, M. (2021), 
Using connectomics for predictive assessment of brain parcellations, 
{\it NeuroImage}, {\bf 238}, 118170 
\url{doi:10.1016/j.neuroimage.2021.118170}

\bibitem{la07}
Lapicque, L. (1907), 
Recherches quantitatives sur l’excitation èlectrique des nerfs traitèe comme
une polarization, 
{\it J. Physiol. Pathol. Gènèrale}, {\bf 9}, 567-578 

\bibitem{br07}
Brunel, N. and van Rossum, M.~C.~W. (2007), 
Quantitative investigations of electrical nerve excitation treated as polarization (translation of “Recherches quantitatives sur l’excitation èlectrique des nerfs traitèe comme une polarization”), 
{\it Biological Cybernetics}, {\bf 97}, 341-349 
\url{doi:10.1007/s00422-007-0189-6} 

\bibitem{ni15} 
Nicosia, V. and Latora, V. (2015), 
Measuring and Modeling Correlations in Multiplex Networks, 
{\it Physical Review E}, {\bf 92}, 032805 
\url{doi:10.1103/physreve.92.032805} 

\bibitem{an22}
Anesiadis, K. and Provata, A. (2022), 
Synchronization in Multiplex Leaky Integrate-and-Fire Networks with Nonlocal Interactions, 
{\it Frontiers in Network Physiology}, {\bf 2}, 910862 
\url{doi:10.3389/fnetp.2022.910862}


\end{thebibliography}
\end{document}